\documentstyle[11pt,twoside,paspconf]{article}
\input{epsf}
\def\kms{\rm ~km~s^{-1}}
\def\etal{{\it et al. }}

\def\gsim{ \lower .75ex \hbox{$\sim$} \llap{\raise .27ex \hbox{$>$}} }
\def\lsim{ \lower .75ex \hbox{$\sim$} \llap{\raise .27ex \hbox{$<$}} }

\def\hmpc{{\rm h}$^{-1}$ {\rm Mpc}\ }
\def\kmsmpc{km s$^{-1}$ Mpc$^{-1}$\ }

\begin{document}

\title{The fate of LSB galaxies in clusters and the origin of the 
diffuse intra-cluster light}

\author{Ben Moore}
\affil{Department of Physics, University of Durham, 
South Road, Durham, DH1 3LE, UK}

\author{George Lake, Joachim Stadel \& Thomas Quinn}
\affil{Department of Astronomy,
University of Washington, 
Seattle, WA98195, USA}

\begin{abstract}

We follow the evolution of disk galaxies within a cluster that forms
hierarchically in a standard cold dark matter N-body simulation.  At a
redshift $z=0.5$ we select several dark matter halos that have quiet
merger histories and are about to enter the newly forming cluster
environment.  The halos are replaced with equilibrium high resolution
model spirals that are constructed to represent luminous examples of
low surface brightness (LSB) and high surface brightness (HSB)
galaxies. Whilst the models have the same total luminosity, $\sim
L_*$, they have very different internal mass profiles, core radii and
disk scale lengths, however they all lie at the same place on the
Tully-Fisher relation.  Due to their ``soft'' central potentials, LSB
galaxies evolve dramatically under the influence of rapid encounters
with substructure and strong tidal shocks from the global cluster
potential \--- galaxy harassment.  As much as 90\% of the LSB disk
stars are tidally stripped and congregate in large diffuse tails that
trace the orbital path of the galaxy and form the diffuse
intra-cluster light.  The bound stellar remnants closely resemble the
dwarf spheroidals (dE's) that populate nearby clusters, with large
scale lengths and low central surface brightness.


\end{abstract}

\section{Introduction}

Clusters of galaxies provide a unique environment wherein the galaxy
population has been observed to rapidly evolve over the past few
billion years (Butcher \& Oemler 1978, Dressler \etal 1998).  At
a redshift $z \gsim 0.4$, clusters are dominated by spiral galaxies
that are predominantly faint irregular or Sc-Sd types. Some of these
spirals have disturbed morphologies; many have high rates of
star-formation (Dressler \etal 1994a).  Conversely, nearby clusters
are almost completely dominated by spheroidal (dSph), lenticulars (S0)
and elliptical galaxies (Bingelli \etal 1987, 1988, Thompson \&
Gregory 1993).  Observations suggest that the elliptical galaxy
population was already in place at much higher redshifts, at which
time the S0 population in clusters is deficient compared to nearby
clusters (Couch \etal 1998, Dressler \etal 1998).  This evolution of
the morphology-density relation appears to be driven by an increase in
the S0 fraction with time and a corresponding decrease in the luminous
spiral population.

Low surface brightness (LSB) galaxies appear to avoid regions of high
galaxy densities (Bothun \etal 1993, Mo \etal 1994).  This is somewhat
puzzling since recent work by Mihos \etal (1997) demonstrated that LSB
disk galaxies are actually {\it more} stable to close tidal encounters
than HSB disk galaxies.  In fact, LSB galaxies have lower disk mass
surface densities and higher mass-to-light ratios, therefore their
disks are less susceptible to internal global instabilities, such as
bar formation.  However, in a galaxy cluster, encounters occur
frequently and very rapidly, on a shorter timescale than investigated
by Mihos \etal and the magnitude of the tidal shocks are potentially
very large.

Several physical mechanisms have been proposed that can strongly
affect the morphological evolution of disks: ram-pressure stripping
(Gunn \& Gott 1978), galaxy merging (Icke 1985, Lavery \& Henry 1988,
1994) and galaxy harassment (Moore \etal 1996a, 1998). The importance
of these mechanisms varies with environment: mergers are frequent in
groups but rare in clusters (Ghigna \etal 1998), ram pressure removal
of gas is inevitable in rich clusters but will not alter disk
morphology (Abadi \etal 1999).  
The morphological transformation in
the dwarf galaxy populations ($M_b>-16$) in clusters since $z=0.4$ can
be explained by rapid gravitational encounters between galaxies and
accreting substructure \-- galaxy harassment.  The impulsive and
resonant tidal heating from rapid fly-by interactions causes a
transformation from disks to spheroidals.

The numerical simulations of Moore \etal focussed on the evolution of
fainter Sc-Sd spirals in static  cluster-like potentials and
their transition into dSph's.  In this work we shall examine the role of
gravitational interactions in driving the evolution of luminous
spirals in dense environments.  We will use more realistic
simulations that follow the formation and growth of a large cluster
that is selected from a cosmological simulation of a closed CDM
universe.  The parameter space for the cluster model is fairly well
constrained once we have adopted hierarchical structure formation.
The structure and substructure of virialised clusters is nearly
independent of the shape and normalisation of the power
spectrum. Clusters that collapse in low Omega universes form earlier,
thus their galaxies have undergone more interactions. The cluster that
we follow virialises at $z\sim 0.3$, leaving about 4 Gyrs for the
cluster galaxies to evolve. 

The parameter space for the model spirals is much larger. Mihos \etal
examined the effects of a single encounter at a fixed number of disk
scale lengths, whilst varying the disk surface brightness and keeping
other properties fixed.  The key parameter that determines whether or
not dark matter halos survive within a cluster N-body simulation is
the core radius of the substructure, which is typically dictated by
the softening length (Moore, Katz \& Lake 1996b).  We suspect that the
``softness'' of the dark matter potentials 
may also be the key factor that governs
whether or not a given disk galaxy will survive within a real
cluster.

\section{The model galaxies}

We use the technique developed by Hernquist (1989) to construct
equilibrium spiral galaxies with disk, bulge and halo components, that
represent ``standard'' HSB and LSB disk galaxies.  In each model the
disk mass is $4.0\times10^{10}M_\odot$ and the rotation curves both
peak at $200\kms$. They are a little less massive than ``$L_*$''
galaxies, the characteristic luminosity of the break in the galaxy
luminosity function and would have absolute magnitudes $\sim -17.8$
for a mass to light ratio of 5.
The ``HSB'' spiral has an exponential disk scale
length, $r_d=3.0$ kpc and a bulge with a mass of one third of the
total disk mass.  The ``LSB'' disk scale length is 10 kpc and has no
bulge.  The scale height, $r_z$, of each disk is $0.1r_d$ and they are
constructed with a Toomre $Q$ parameter of 1.5.  Each galaxy has a
dark halo modeled by truncated isothermal spheres with core radii set
equal to the disk scale length. This scaling ensures that each galaxy
lies at the same point on the Tully-Fisher relation, yet the galaxies
will have different internal mass distributions (Zwaan \etal 1995).  

\centerline{\epsfxsize=\hsize\epsfbox{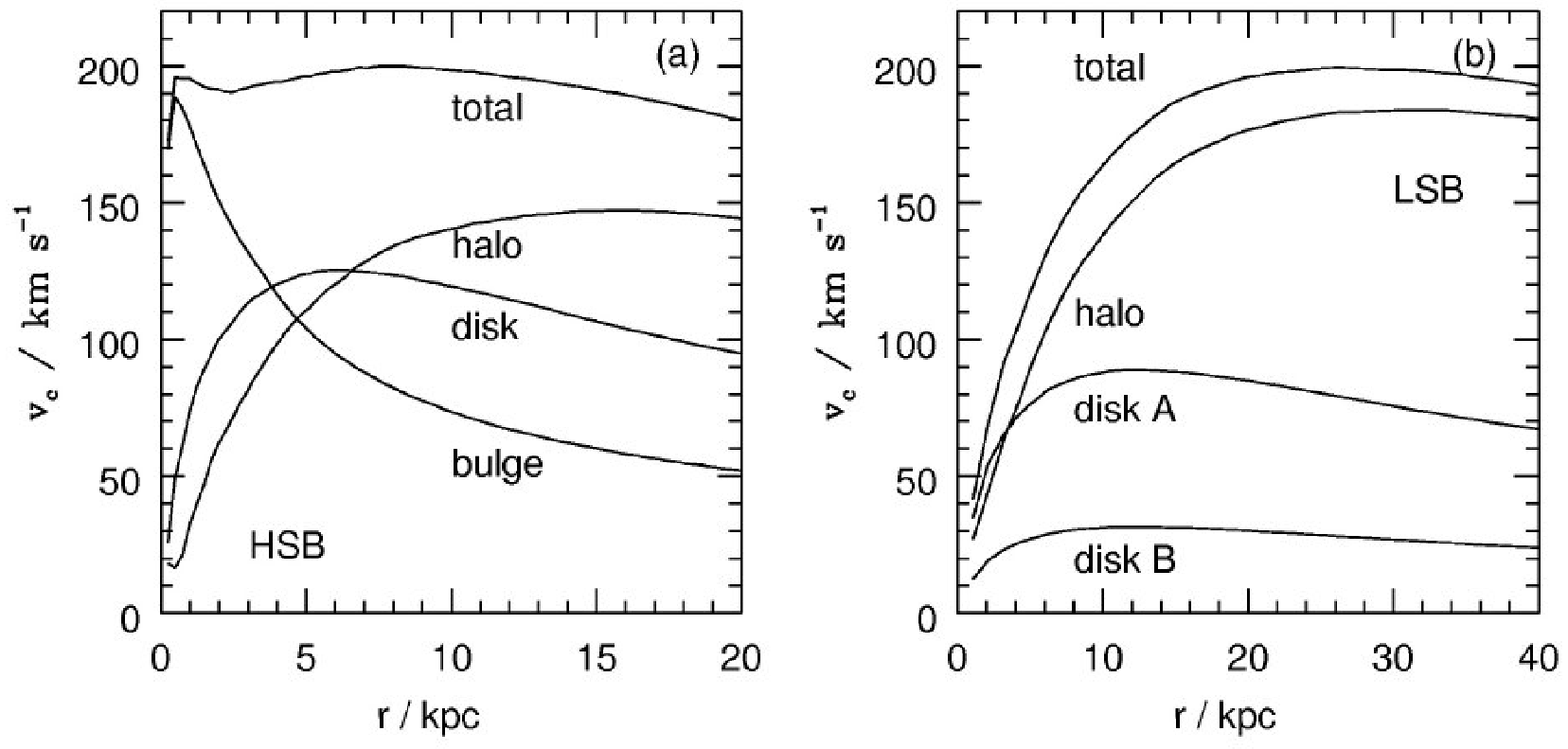}}
{\it Figure 1. The curves show the contributions from
stars and dark matter to the total rotational velocity of the disk
within (a) the HSB galaxy and (b) the LSB galaxy.}

\

Figure 1 shows the contribution to the rotation velocity of the disks
from each component.  Note that the bulge component of the HSB galaxy
has ensured that the rotation curve is close to flat over the inner 5
disk scale lengths, whereas the rotation curve of the LSB galaxy rises
slowly over this region.  These rotation curves are typical of that
measured for LSB galaxies (de Blok \& McGaugh 1996) and HSB galaxies
(Persic \& Salucci 1997). 

Each disk is modeled using 20,000 star particles of mass
$2\times10^6M_\odot$ and 40,000 dark matter halo particles of mass
$2\times10^7M_\odot$ in the LSB and $6\times10^6M_\odot$ in the HSB
galaxy.  The force softening is $0.1r_d$ for the star particles and
$0.5r_d$ for the halo particles. Their disks are stable and they
remain in equilibrium when simulated in isolation. Discreteness in the
halo particles causes the disk scale height to increase with time as
quantified in Section 4 for the LSB galaxy.  


\section{The response to impulsive encounters}

For a given orbit through a cluster, the visible response of a disk
galaxy to a tidal encounter depends primarily upon its internal
dynamical timescale.  Galaxies with cuspy central mass distributions,
such as ellipticals, have short orbital timescales at their centres
and they will respond adiabatically to tidal perturbations.  Sa-Sb
spirals have flat rotation curves, therefore a tidal encounter will
cause an impulsive disturbance to a distance $\sim v_c b/V$, where $b$
is the impact parameter, $V$ is the encounter velocity and $v_c$ is
the galaxy's rotation speed. LSB galaxies and Sc-Sd galaxies have
slowly rising rotation curves, indicating that the central regions are
close to a uniform density.  The central dynamical timescales are
constant throughout the inner disk and an encounter that is impulsive
at the core radius will be impulsive throughout the galaxy.

\

The strength of an encounter is $\propto M_p^2/V^2$, where $M_p$ is
the perturbing mass. The typical galaxy-galaxy encounter within a
virialised cluster occurs at a relative velocity $\sim \sqrt
2\sigma_{1d}$. Substituting typical parameters for an Sa--Sb spiral
orbiting within a cluster, we find that such encounters will not
perturb the disk within $\sim 3r_d$.  However, tidal shocks from the
mean cluster field also provides a significant heating source for those
galaxies on eccentric orbits (Byrd \& Valtonen 1990, Valluri 1993).
Ghigna \etal (1998) studied the orbits of several hundred dark halos
within a cluster that formed hierarchically in a cold dark matter
universe. The median ratio of apocenter to pericenter was 6:1, with a
distribution skewed towards radial orbits.  More than 20\% of the
halos were on orbits more radial than 10:1. A galaxy on this orbit would
move past pericenter at several thousand $\kms$ and would
be heated across the entire disk.

\

We illustrate the effect of a single impulsive encounter on each of
our model disks in Figure 2 and Figure 3.  At time t=0 we send a
perturbing halo of mass $2\times 10^{12}M_\odot$ perpendicular to the
plane of the disk at an impact parameter of 60 kpc and velocity of
1500 $\kms$.  This encounter would be typical of that occurring in a
rich cluster with a tidally truncated $L_*$ elliptical galaxy near the
cluster core.  Any one galaxy in the cluster will suffer several
encounters stronger than this since the cluster formed.  Although we
simulate a perpendicular orbit here, we do not expect the encounter
geometry to make a significant difference since the difference between
direct and retrograde encounters will be small {\it i.e.} $V >> v_c$.

\

At t=0.1 Gyrs after the encounter, the perturber has moved 150 kpc
away, yet the visible response to the encounter is hardly
apparent. After 0.2 Gyrs, we can begin to see the response to the
tidal shock as material is torn from the disk into extended tidal
arms.  Even at this epoch their is a clear difference to the response
of the perturbation by each galaxy.  After 0.4 Gyrs, the LSB galaxy is
dramatically altered over the entire disk and a substantial fraction
of material has been removed past the tidal radius.  Remarkably, 
the central disk of the HSB galaxy remains intact and only the
outermost stars have been strongly perturbed.

\centerline{\epsfysize=3.3truein \epsfbox{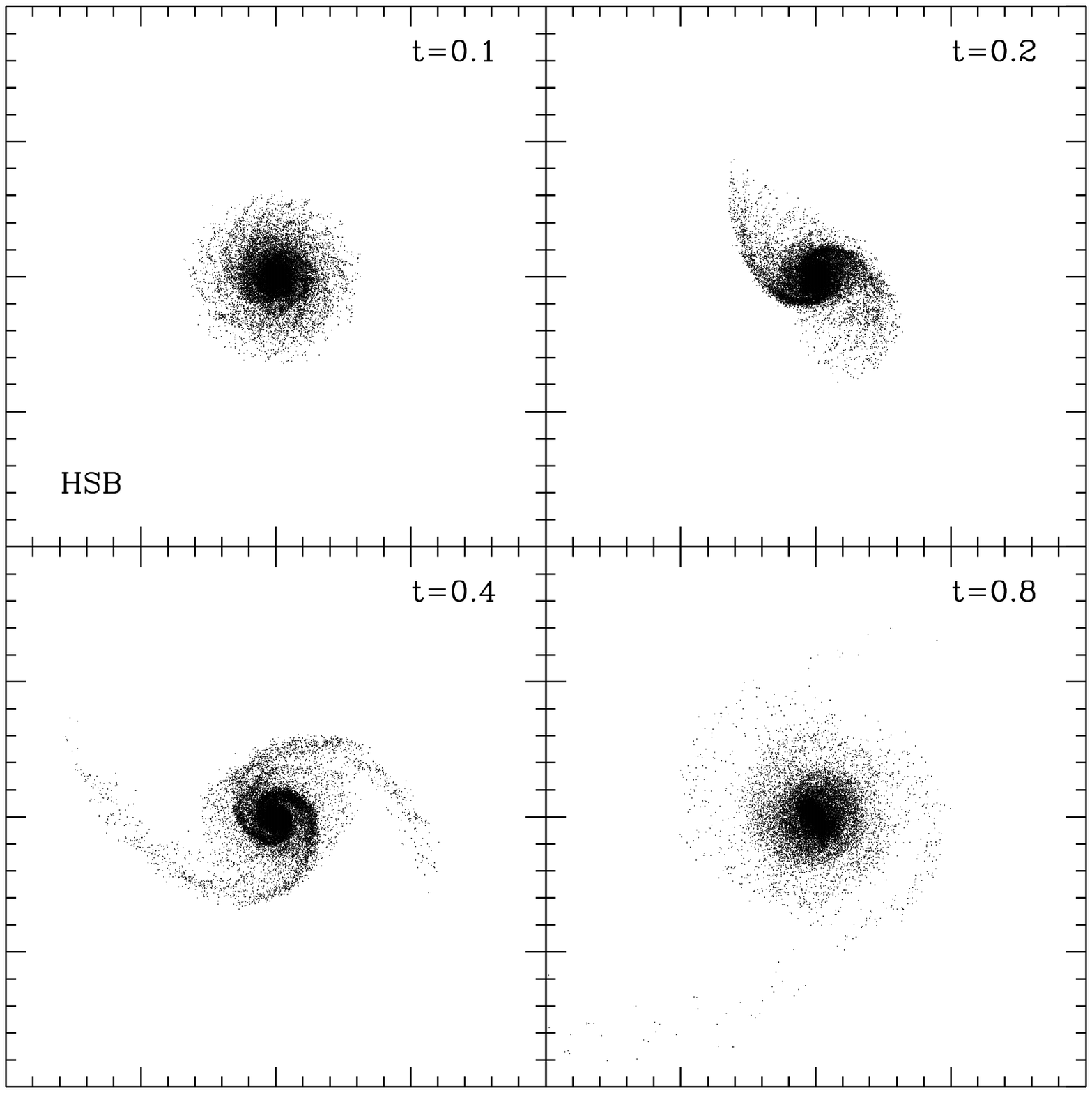}}

{\it Figure 3. Snapshots of the distribution of disk stars from a HSB
galaxy after a single high-speed encounter with a massive galaxy.
Each frame is 120 kpc on a side and encounter takes place
perpendicular to the disk at the box edge (60 kpc).
}

\centerline{\epsfysize=3.3truein \epsfbox{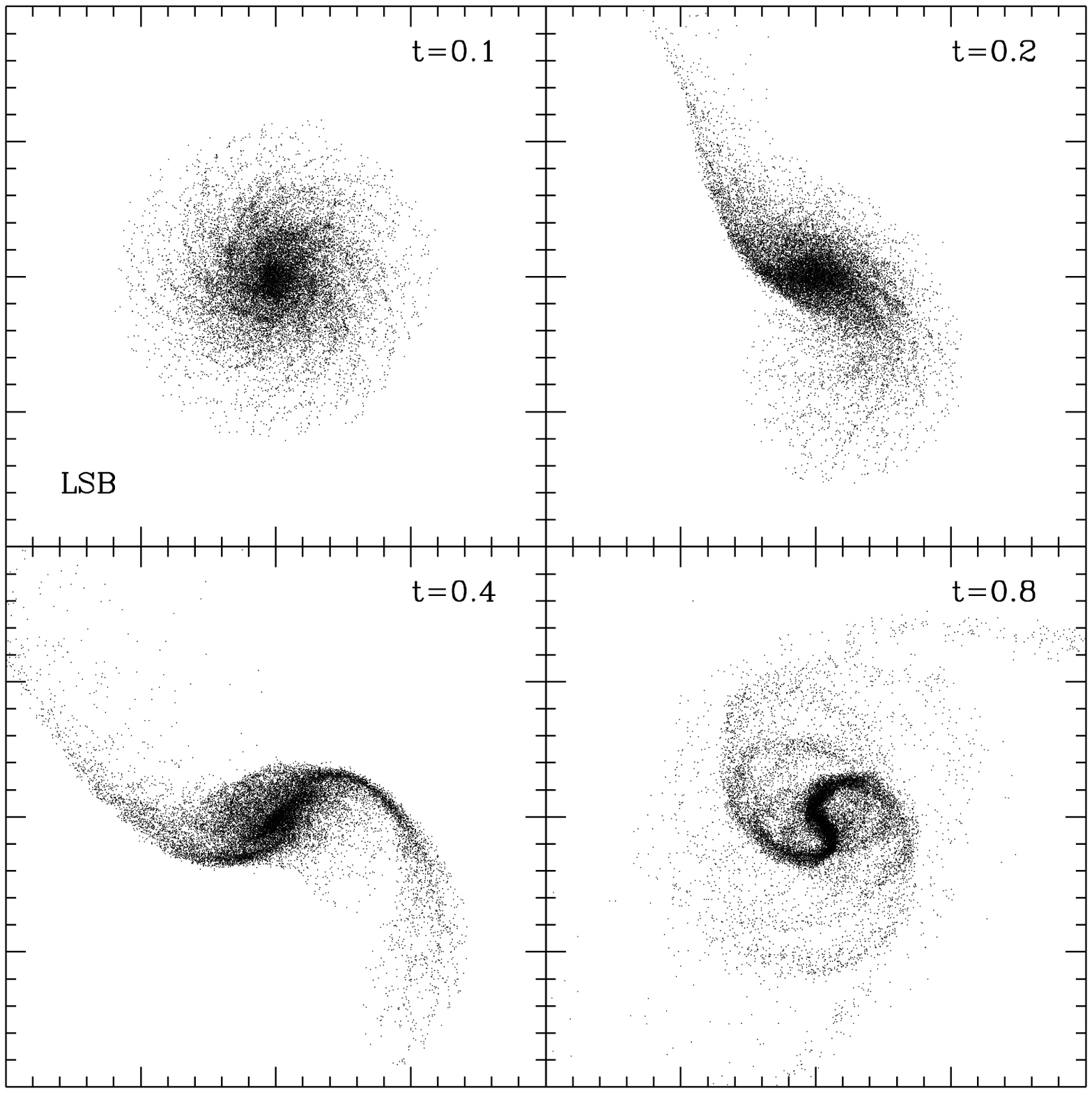}}

{\it Figure 4. Snapshots of the distribution of disk stars from an LSB galaxy
after a single high-speed encounter with a massive galaxy.  
Each frame is 120 kpc on a side and the encounter takes place
perpendicular to the disk at the box edge (60 kpc).}

\section{Simulating disk evolution within a hierarchical universe}

Previous simulations of tidal shocks and galaxy harassment focussed
upon the evolution of disk galaxies in static clusters with
substructure represented by softened potentials with masses drawn from
a Schechter function (Moore \etal 1996a \& 1998).  Here we use a more
realistic approach of treating the perturbations by following the
growth of a cluster within a hierarchical cosmological model. The
cluster was extracted from a large CDM simulation of a closed universe
within a $50$ Mpc box and was chosen to be virialised by the present
epoch.  (We assume a Hubble constant of 100 \kmsmpc \ .)  Within the
turn-around region there are $\sim 10^5$ CDM particles of mass
$10^{10}M_\odot$ and their softening length is 20 kpc.  At a redshift
$z=0$ the cluster has a one dimensional velocity dispersion of
$700\kms$ and a virial radius of 2\hmpc. The tidal field from the mass
distribution beyond the cluster's turn-around radius is simulated with
massive particles to speed the computation.


Between a redshift z=2 to z=0.5 we follow the merger histories of
several candidate dark matter halos from the cosmological simulation
that end up within the cluster at later times.  We select three halos
with circular velocities $\sim 200\kms$ that have suffered very little
merging over this period and would therefore be most likely to host
disk galaxies. We extract these halos from the simulation at z=0.5 and
replace the entire halo with the pre-built high resolution model
galaxies.  We rescale the disk and halo scale lengths by $(1+z)^{-1}$
according to the prescription of Mao \etal (1998) to represent the
galaxies entering the cluster at higher redshifts.  On a 32 node
parallel computer, each run takes several hours; three runs were
performed in which the halos were replaced with LSB disks and a
further three runs using HSB disks.

\centerline{\epsfysize=2.6truein \epsfbox{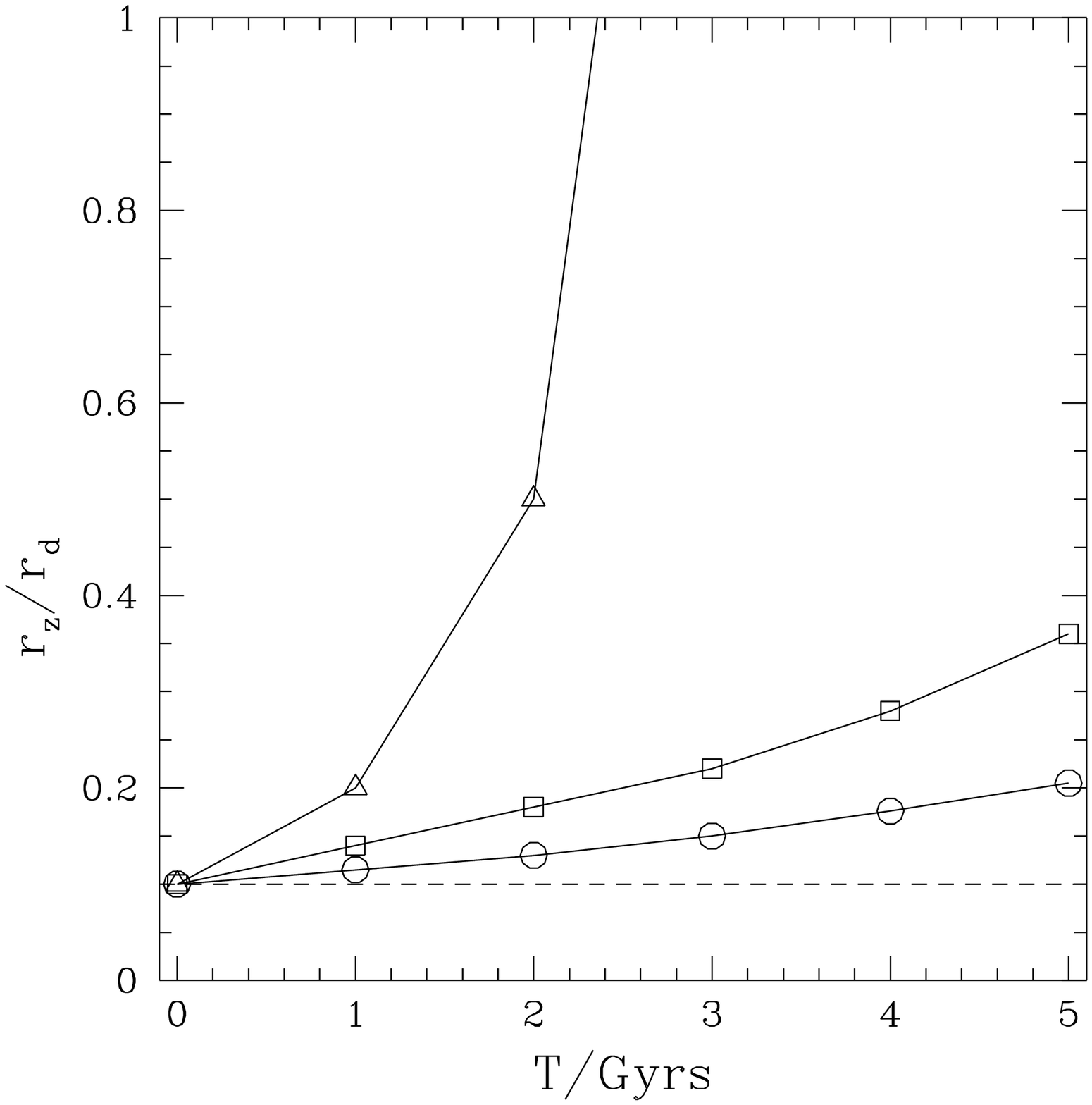}}

{\it Figure 4. The vertical scale height, $r_z$, of the 
disk in units of the initial 
disk scale length, $r_d$, measure at $r_d$ and plotted against time. 
The circles show the HSB galaxy placed in a void to test the 
numerical heating. The squares 
and triangles show one of the HSB and LSB galaxies that enters
the cluster respectively.}

\

At z=0.5, the cluster is only just starting to form from a series of
mergers of several individual group and galaxy sized halos. The
cluster quickly virialises, although several dark matter clumps
survive the collapse and remain intact orbiting within the clusters
virial radius.  Between $z=0.4\--0.3$ the model galaxy receives a
series of large tidal shocks from the halos that are assembling
the cluster. Once the galaxy enters the virialised cluster, it
continues to suffer encounters with infalling and orbiting
substructure.  By a redshift z=0.1, many stars have been
stripped from the disk and now orbit through the cluster - closely
following the rosette orbit of the parent galaxy.  Of the three LSB
galaxy runs, between 60\% and 90\% of the stars were harassed from the
disk, whereas the stellar mass loss in the HSB runs was between
10\%--30\%.

\section{Summary}

The response of a disk galaxy to tidal shocks is governed primarily by
the concentration of the mass distribution that encompasses the
visible disk. LSB galaxies have slowly rising rotation curves and
dynamical timescales that are constant within their central regions.
LSB galaxies cannot survive the chaos of cluster formation;
gravitational tidal shocks from the merging substructure literally
tear these systems apart, leaving their stars orbiting freely within
the cluster and providing the origin of the intra-cluster light
({\it c.f.} Moore \etal 1999).

Recent observations of individual planetary nebulae within clusters,
but outside of galaxies, lends support to this scenario.  Estimates of
the total diffuse light within clusters, using CCD photometry
(Bernstein \etal 1995, Tyson \& Fischer 1995) or the statistics of
intra-cluster stars (Theuns \& Warren 1997, Feldmeier \etal 1998,
Mendez \etal 1998, Ferguson \etal 1998), ranges from 10\% to 45\% of
the light attached to galaxies. Presumably, these stars must have
originated within galactic systems. The integrated light within LSB
galaxies may be equivalent to the light within ``normal'' spirals
(Bothun, Impey \& McGaugh 1997, and references within).  This is
consistent with the entire diffuse light in clusters originating from
harassed LSB galaxies.


High surface brightness disk galaxies and galaxies with luminous
bulges have steep mass profiles that give rise to flat rotation curves
over their visible extent.  The orbital time within a couple of disk
scale lengths is short enough for the disk to respond adiabatically to
rapid encounters.  Tidal shocks cannot remove a large amount of
material from these galaxies, nor transform them between morphological
types, but will heat the disks and drive instabilities that can funnel
gas into the central regions (Lake \etal 1998).  A few Gyrs after
entering a cluster, their disks are thickened and no spiral features
remain.  If ram-pressure is efficient at removing gas from disks, we
speculate that these galaxies will rapidly evolve into SO's.  Since
the harassment process and ram-pressure stripping are both more
effective near the cluster centers, we expect that a combination of
these effects may drive the morphology--density relation within
clusters.


\end{document}